\documentclass[aps,prb,epsfig,twocolumn]{revtex4}
\usepackage{amssymb,amsmath}

\begin{document}

\draft

\title{ Darwinian Dynamics Implies Developmental Ascendency }

\author{ Ping Ao }
\address{ Department of Mechanical Engineering and Department of Physics \\
               University of Washington, Seattle, WA 98195, USA  }

\date{Nov. 19 (2006); Aug. 7 (2007) }

\begin{abstract}
 A tendency in biological theorizing is to formulate principles above or equal
 to Evolution by Variation and Selection of Darwin and Wallace.
%
%
In this letter I analyze one such recent proposal which did so for
the developmental ascendency. I show that though the idea of
developmental ascendency is brilliant, this is in wrong order in
the hierarchical structure of biological theories and can easily
generate confusing. Several other examples are also briefly
discussed in the note added.
\\ {Published version: {\bf Biological Theory 2}(1) (2007) 113-115. }
\end{abstract}


\maketitle


In a thoughtful essay Coffman (2006) argued for the need of a
general theoretical framework for developmental processes.
He proposed that such a general theory, which he named the
developmental ascendency (DA), indeed exist. He further argued
that such a theory may supersede Darwinian evolutionary theory.
While his first part of argument appears convincing, his
contention that DA is superior to Darwinian evolution are in wrong
theoretical order. It has been demonstrated recently (Ao 2005),
though the main ideas were already in earlier seminar works of
Fisher (1930) and Wright (1932), that Darwinian evolution implies
thermodynamics, therefore DA. To the present author, such a
mixed-up is not accidental: It reflects a conceptual defect in the
understanding of Darwinian evolution persistent to these days ever
since neo-Darwinism. The purpose of the present letter is to point
out and to clarify this misunderstanding.

There are two profound and quantitative concepts embedded in
Darwinian evolutionary theory (Fisher 1930; Wright 1932; Ao 2005).
The first is Fisher's fundamental theorem of natural selection
(FTNS), which links the variation and stochasticity to the ability
of a system seeking best places in its enormous functional space
during its evolution. This intrinsic dynamical concept was
initially formulated within the context of population genetics
more than 70 years ago. Since then, its role has been found in all
biological processes in which developmental processes are special
cases. Its counter-part in physical sciences is the so-called
fluctuation-dissipation theorem (Ma 1985), one of the most
important theorems in modern physics, associated with great names
such as Einstein. The second fundamental concept is equally
important, known as Wright's adaptive landscape, again proposed
more than 70 years ago.  It describes the possible final selection
to which the system to evolve. Its physical science counter-part
is the potential energy or Hamiltonian (Ao 2005), the very
quantity needed in any dynamical discussion in physics. Following
Darwinian evolutionary theory, with the aid of FTNS and adaptive
landscape, it has been demonstrated mathematically that
thermodynamics is determined by Wright potential function in
Darwinian dynamics (Ao 2005, 2006). As argued by Coffman, DA is
implied in the thermodynamics. Therefore, these two logic
relations imply that DA is a special case of Darwinian
evolutionary theory. This would have already been suggested by the
very terminologies associated with those general theories:
Thermodynamics means no dynamics, because it only concerns with
equilibrium or steady state processes; On the other hand,
Darwinian evolutionary theory is an explicit nonequilibrium
dynamical theory. The description of the mathematical structure
embedded in its final steady state is thermodynamics.

Logically I have now reached the end of my reasoning: It have been
demonstrated that Darwinian evolutionary dynamics implies DA, not
the other way. Nevertheless, it is not so simple in reality. Let
me further address a few more specific and interesting issues
raised in Coffman's essay to make my point more clear.

{\bf Acceptance of adaptive landscape in biology.} In the Abstract
his essay Coffman acknowledged that the general idea of his DA was
already implied in a somewhat ignored Ulanowicz's ascendency
theory in ecology (Ulanowicz 1980). In text he discussed a few
reasons for such neglecting, which are very plausible. Here I
would add two more reasons. The first lies in the mathematics.
Until recent, there was no general mathematical framework to
accommodate Wright's adaptive landscape in a generic
nonequilibirum setting. A detailed mathematical discussion is
beyond the present letter.  Relevant literature can be found
elsewhere (Ao 2005, 2006).

The second additional reason is biological. Ever since the
formulation of neo-Darwinism, Fisher's FTNS and Wright's adaptive
landscape had not been fully understood and had been
controversial. Those misunderstandings had been summarized by
Gould (2002), a book cited in favor in Coffman's essay. While
Gould's criticism on neo-Darwinism may be valid, he had no good
reason to dismiss those two quantitative concepts as violating
contingency, individuality, and interaction, three ideas central
in Darwinian evolutionary theory. It has been shown that after
clarifying the imprecise statements associated with the original
formulations, Fisher's FTNS and Wright's adaptive landscape are
indeed consistent with those three central ideas in evolution (Ao
2005). Furthermore, Gould's dismissing appears rather odd viewing
from the punctuated equilibrium theory which Gould himself helped
to establish: It is now known that Fisher's FTNS and Wright's
adaptive landscape actually provide a sound theoretical and
quantitative foundation to discuss punctuated equilibrium.

If such confusions were not clarified, it would be likely that
Coffman's DA would suffer the same fate as that of Ulanowicz's.

{\bf Natural selection vs autocatalytic cycle.} Coffman suggested
that "natural selection is an emergent property of autocatalytic
cycles".  Again, this is in wrong order. First of all, it has been
universally accepted that at its beginning there did not exist
such cycles on Earth. A complete discussion of this issue would go
into philosophical issues regarding to teleological reasoning
beyond my letter. Suffice it to say that natural selection gives
arise to the autocatalytic cycles and the variation is the source
of innovation and creation. Without variation, we would be in a
deterministic world, a perspective rightly argued against by
Coffman. Once such cycles are fixed by selection, or, borrowing
the language from Waddington (1959), are canalized,
it provides the adaptive landscape for further evolution and,
concomitantly, for further selection.

Such dynamical phenomena suggest that Darwinian evolutionary
theory has essentially two parts: the dynamical structure encoded
in evolution by selection and variation, further classified into
four dynamical elements by following ideas of Fisher and Wright
(Ao 2005, 2006);
and the structure of each dynamical element in a given biological
process. Same classification of two types of fundamental laws in
biology has also been proposed recently from a completely
different perspective (Wilson 2006). Today it should be evident
that Darwin and Wallace (Darwin 1958; Darwin and Wallace 1858) got
the dynamical part completely correct nearly 150 years ago. Since
then, this dynamical part has been demonstrated to be independent
of specific biological instantiation and has been applied to
biological processes ranging from gene regulation, metabolic
pathways, ecology, cognitive science, etc (Wilson 2006). This
dynamical part of evolutionary theory is sufficient to give arise
to the developmental ascendency as reasoned above. Nevertheless,
it is also well known that Darwin and Wallace did not get the
structures of dynamical elements for the heretic process correct,
among others: Mendel's theory was unknown to them. Clearly, we too
have not completely understood all structures of all dynamical
elements, such as those in developmental processes. The desire to
understand such structures has been one of major driving forces in
modern biological and medical research.

Coffman pointed out that the gene regulation is an example to
illustrate his DA, with both positive and negative feedbacks,
abundant in developmental processes (Davidson 2006),
with which I completely agree. I wish to further point out that
Darwinian dynamics exemplified by adaptive landscape and
stochastic force has been quantitatively used in the modelling a
gene regulatory network, the lambda genetic switch formed by both
positive and negative feedbacks (Ptashne 2004), where the
dynamical elements are determined by the Jacob-Monod operon theory
constraint by physical and chemical laws. It has now achieved
quantitative agreement between the theory and experiments and with
further predictions (Zhu {\it et al} 2004). There has been a
continuous progress in the experimental determination of the gene
regulatory structure in the complex developmental processes such
as that of sea urchins (Davidson 2006). It is conceivable that a
biological theory similar to that of Jacob-Monod to specify the
dynamical elements may soon be available and that a quantitative
and predicative Darwinian dynamics study of such complex
developmental processes may be readily carried out in various
theoretical/computational labs. This may be regarded as an
explicit example for the universality of Darwinian dynamics.

{\bf Darwin and Einstein.} Coffman correctly observed that
"Einstein's general theory of relativity contextualizes Newtonian
physics". However, above analysis shows that the same logic cannot
be applied to DA with regard to Darwinian evolutionary theory.
Einstein had deeply modified the dynamical structure of Newtonian
dynamics. DA has not changed, and will not change in my opinion,
the structure of Darwinian dynamics. The four dynamical elements
of Darwinian dynamics has already appeared to be what needed in
the mathematical modelling of developmental processes. Darwin and
Wallace's word equation,
\begin{center}
  Evolution by Variation and Selection,
\end{center}
\noindent remains its supreme position in biology. It may even
help physicists to solve some of their own difficult problems, as
having been trying. Darwinian dynamics implies developmental
ascendency, not the other way.

{\ }

{\ }

\noindent {\it Note added, Aug. 6 (2007).} An interesting article
in New York Times a week ago \cite{nyt2007} promoted the
cooperation theory (CT), where CT was placed as equal to the
Evolution by Variation and Selection. Given above analysis, this
still appears a little uneven: CT is only one of many realizations
of Darwinian dynamics. Let me elaborate this point further below,
because it strengthens the argument in the text.

As already discussed \cite{ao2005,wilson}, there are two types
fundamental laws. One is of course the dynamical laws embedded in
Evolution by Variation and Selection. The other is its
realization, which is arguably best represented by the theory of
Mendel-Watson-Crick, at least for what we have known so far on
Earth. The advancement in human understanding of Nature may
generate (artificial) life not bounded by DNA and RNA in the
future. For many biological studies, the Mendel-Watson-Crick
theory is too fundamental to be convenient. More effective
theories at given phenomenological levels should be, and have
been, developed. The theory of developmental ascendency
\cite{coffman} is such an example. The well known
self-organization theory (SOT) is another realization \cite{sole},
and CT appears to a special case of SOT. Let's grant CT's
independence from SOT to simplify the presentation. Both CT and
SOT have broad explanatory power. For example, they may explain
the origin of religions, and,  both CT and SOT, and possibly
others, are "everywhere in evolution where interesting things are
happening". Still, CT is one realization of Evolution by Variation
and Selection, not "one of the three basic principles of
evolution. The other two are mutation and selection." The upshot
is that, search for appropriate effective biological theories in
the framework of Darwin and Wallace was one of the central tasks
in last century. It will remain so in this century.

\end{document}